\documentclass[11pt,3p]{elsarticle2}
\usepackage[utf8]{inputenc}

\usepackage{caption}
\usepackage{subcaption}
\usepackage{multirow}

\usepackage{graphicx}
\usepackage{amsmath}
\usepackage{amssymb}
\usepackage{amsfonts}
\usepackage{url}
\usepackage{hyperref}
\usepackage[english]{babel}
\usepackage{color}
\usepackage{xspace}

\newcommand{\alphas}{\alpha_{\rm s}}

\newcommand{\sqrts}{\sqrt{\rm s}}

\newcommand{\epem}{e^+e^-}

\providecommand{\ffbar}{f\overline{f}}
\providecommand{\qqbar}{q\overline{q}}
\providecommand{\bbbar}{b\overline{b}}
\providecommand{\uubar}{u\overline{u}}
\providecommand{\ddbar}{d\overline{d}}
\providecommand{\ssbar}{s\overline{s}}
\providecommand{\ccbar}{c\overline{c}}

\newcommand{\AFBb}{A_{_{\textsc{fb}}}^{0,b}}
\newcommand{\AFBbb}{A_{_{\textsc{fb}}}^{b}}
\newcommand{\AFBobs}{(A_{_{\textsc{fb}}}^{b})_{_{\rm obs}}}
\newcommand{\weakang}{\sin^2\theta_{\rm W}}
\newcommand{\weakeff}{\sin^2\theta_{_{\rm eff}}^f}
\newcommand{\weakeffb}{\sin^2\theta_{_{\rm eff}}^b}
\newcommand{\pythia}{{\sc pythia}}
\newcommand{\vincia}{{\sc vincia}}
\newcommand{\jetset}{{\sc pythia\,5/jetset}}
\newcommand{\jade}{{\sc jade}}
\newcommand{\fastjet}{{\sc FastJet}}
\newcommand{\herwig}{{\sc Herwig}}

\newcommand*{\eg}{e.g.\@\xspace}
\newcommand*{\ie}{i.e.\@\xspace}

\def\mean#1{\ensuremath{\left<#1\right>}}

\begin{document}

\begin{frontmatter}

\title{Revised QCD effects on the Z\,$\to\,\bbbar$ forward-backward\\ asymmetry in $\epem$ collisions}
\author{David d'Enterria}
\ead{david.d'enterria@cern.ch}
\address{CERN, EP Department, CH-1211 Geneva 23, Switzerland}
\author{Cynthia Yan}
\ead{cyan2019@stanford.edu}
\address{Stanford Institute for Theoretical Physics, Stanford University, Stanford CA 94305 USA}

\begin{abstract}
The forward-backward (FB) asymmetry of $b$ quarks in $\epem$ collisions at the Z pole measured at LEP, $\AFBb = 0.0992\pm0.0016$, remains today one of the electroweak precision observables with the largest disagreement (2.4$\sigma$) with respect to the Standard Model prediction, $(\AFBb)_{_{\rm th}} = 0.1030 \pm 0.0002$. Beyond the dominant statistical uncertainties, QCD effects, such as $b$-quark showering and hadronization, are the leading sources of $\AFBb$ systematic uncertainty, and have not been revised in the last twenty years. We reassess the QCD uncertainties of the eight original $\AFBb$ LEP measurements, using modern parton shower \pythia\,8 and \vincia\ simulations with nine different implementations of soft and collinear radiation as well as of parton fragmentation. Our analysis, combined with NNLO massive $b$-quark corrections independently computed, indicates total propagated QCD uncertainties of $\sim$0.7\% and $\sim$0.3\% for the lepton- and jet-charge analyses, respectively, that are about a factor of two smaller than those of the original LEP results. Accounting for such updated QCD effects leads to a new $\AFBb = 0.0995\pm0.0016$ average, with a data-theory tension slightly reduced from 2.4$\sigma$ to 2.2$\sigma$. Confirmation or resolution of this long-term discrepancy requires a new high-luminosity $\epem$ collider collecting orders-of-magnitude more data at the Z pole to significantly reduce the dominant $\AFBb$ statistical uncertainties, and to improve our understanding of $b$-quark showering and hadronization.
\end{abstract}


\end{frontmatter}

%%%%%%%%%%%%%%%%%%%%%%%%%%%%%%%%%%%%%%%%%%%%%
\section{Introduction}

In the Standard Model (SM), the fermions are arranged in weak-isospin doublets for left-handed particles and weak-isospin singlets for right-handed particles. The Z boson mediates weak neutral current interactions between fermions, %of the same generation, %, with mixed weak-isospin and electromagnetic couplings. 
and it couples to both left- and right-handed fermions with different strengths depending on their weak-isospin $I^f$ and electric charge $Q^f$. The vector and axial-vector Z couplings for a fermion of type $f$ are $g^f_V = (g_L^f + g_R^f) = I_3^f - 2Q^f\weakang$ and $g^f_A = (g_L^f -g_R^f) = I_3^f$ respectively, where $I_3^f$ is the third component of the weak isospin of the fermion, related to the electric charge via its hypercharge $Y^f$: $Q^f=I_3^f+Y^f/2$, and where $\weakang$ is the weak mixing angle that controls the $\gamma$--Z mixing and connects the coupling constants of the electroweak theory: $g\sin\theta_W=g'\cos\theta_W=e$. For $m_W = 80.3692$~GeV and $m_Z = 91.1880$~GeV, one has $\weakang \equiv 1 - m_W^2/m_Z^2 = 0.22321$ in the on-shell electroweak scheme~\cite{PDG}. The $g^f_{V,A}$ expressions above describe the varying strengths of the Z-fermion couplings in the $(\nu_e,\nu_\mu,\nu_\tau)$, $(e,\mu,\tau)$, $(u,c,t)$, and $(d,s,b)$ lepton/quark families. The mixed vector and axial-vector Z couplings not only affect the total $\epem\to\ffbar$ cross section $\sigma^{\rm tot}_{\ffbar}$, but also induce asymmetries in the angular distributions of the final-state fermions produced. 
%$\epem\to\ffbar$ -- beyond those issuing from the incoming $e^\pm$ helicity, and from the polarisation of the produced particles -- 
%\frac{}{}=\frac{}{}\left[(1-\mathcal{P}_e\mathcal{A}_e)(1+cos^2\theta)+\right]
The Born-level differential cross section for Z exchange alone, as a function of the scattering angle $\theta$ of the outgoing fermion with respect to the direction of the incoming $e^-$ beam, summed over final-state helicities and assuming unpolarized $\epem$  beams, is:
$\frac{\mathrm{d}\sigma_{_{\ffbar}}}{\mathrm{d}\cos\theta}=\frac{3}{8}\sigma^{\rm tot}_{_{\ffbar}}\left((1+\cos^2\theta)+2\mathcal{A}_f\mathcal{A}_e\cos\theta\right)$.
The angular asymmetry parameters $\mathcal{A}_f$ in the $\epem\to\ffbar$ final states %, ignoring beam polarizations, 
are thereby ultimately driven by $Q^f$ and $\weakang$:
\begin{eqnarray}
\mathcal{A}_f=\frac{(g_{L}^f)^2-(g_{R}^f)^2}{(g_{L}^f)^2+(g_{R}^f)^2}=2\,\frac{g_{V}^f/g_{A}^f}{1+(g_{V}^f/g_{A}^f)^2}\,,
%\mbox{ with }\;\;\frac{g_{V}^f}{g_{A}^f}=1-\frac{2Q_f}{I_3^f}\sin^2\theta^f_{\text{eff}}=1-4|Q_f|\sin^2\theta^f_{\text{eff}}
\mbox{ with }\;\;\frac{g_{V}^f}{g_{A}^f}=1-4|Q^f|\sin^2\theta^f_{_{\rm eff}}\,,
\label{eq:Af_vs_Q_sin2Weff}
\end{eqnarray}
where $\weakeff$ is the effective fermion mixing angle that includes electroweak corrections beyond the tree level: $\weakeff = \kappa_f\weakang$, with $\kappa_f$  a flavor-dependent scaling factor absorbing the higher order corrections. The $\epem\to\ffbar$ forward-backward (FB) asymmetry can be determined from the ratio of the difference over the sum of the number of forward $N_F$ and backward $N_B$ events, where forward (backward) indicates that the outgoing fermion (antifermion) is produced in the hemisphere defined by the direction of the electron (positron) beam:
\begin{eqnarray}
A_{\rm FB}^f=\frac{N_F-N_B}{N_F+N_B},\;\;\mbox{ where }\;N_F=\int_0^1 \frac{\mathrm{d}N}{\mathrm{d}\Omega}d\Omega,\;\;N_B=\int_{-1}^0 \frac{\mathrm{d}N}{\mathrm{d}\Omega}d\Omega\,.
\label{eq:AFB_N}
\end{eqnarray}
Experimentally, the FB asymmetries are usually derived from fits of the differential distribution of events as a function of the polar angle $\theta$ between the e$^\pm$ and outgoing fermion directions, which is a procedure statistically slightly more powerful than the simple event counting given by Eq.~(\ref{eq:AFB_N}). From the measured FB asymmetries of the produced $f$ and $\overline{f}$ fermions fitted at various $\epem$ center-of-mass energies ($\sqrts$) around the Z resonance, one derives the relevant asymmetry parameter at the Z pole, $A_{_{\textsc{fb}}}^{0,f}$.

Among the FB quark asymmetries in $\epem\to \mathrm{Z}\to\qqbar(g)$ at $\sqrts \approx m_{\rm Z}$, the $b$-quark one ($\AFBbb$) is the most accurately measured at LEP, with a 1.6\% precision~\cite{ALEPH:2005ab}. This is so because $b$-quark jets are the easiest ones to be properly identified (through the secondary vertex associated with the decays of their leading $B$ hadrons) and to be experimentally assigned a charge, as well as because the large Z coupling to down-type quarks leads to a $\sim$30\% larger branching fraction for Z\,$\to \ddbar,\ssbar,\bbbar$ ($\mathcal{B} = 15.6\%$) compared to Z\,$\to \uubar,\ccbar$ ($\mathcal{B} = 11.6\%$)~\cite{PDG}. The value $\AFBb = 0.0992\pm0.0016$, obtained from the combination of eight measurements at $\sqrts = 91.21$--91.26~GeV using two different (lepton- and jet-charge based) methods~\cite{ALEPH:2005ab}, shows today one of the largest tensions ($2.4\sigma$) with respect to the theoretical SM prediction, $(\AFBb)_{_{\rm th}} = 0.1030 \pm 0.0002$, derived from the combined fit to all precision electroweak observables~\cite{PDG}. Such a discrepancy also propagates to the effective weak mixing angle derived from it, via Eq.~(\ref{eq:Af_vs_Q_sin2Weff}), $\weakeff = 0.23221\pm0.00029$, to be compared with the $\weakeff = 0.23129\pm0.00004$ world-average at the Z pole~\cite{PDG}, in the $\overline{\mathrm{MS}}$ scheme.

The experimental uncertainties of the individual $\AFBb$ extractions at LEP include 3--11\% (1--6\%) statistical (systematic) components~\cite{ALEPH:2005ab}. Among the systematic uncertainties, those that have an origin in quantum chromodynamics (QCD) effects, such as those related to angular decorrelations induced in the event axis due to hard and/or soft and collinear parton radiation and/or parton hadronization, were estimated more than twenty years ago by combining next-to-leading-order (NLO) perturbative corrections with Monte Carlo (MC) parton shower simulations~\cite{Abbaneo:1998xt}. A more recent theoretical study~\cite{Bernreuther:2016ccf} has reduced the perturbative QCD (pQCD) uncertainties by calculating next-to-next-to-leading order (NNLO) corrections with non-zero $b$-quark mass, but the impact of parton shower (PS) and fragmentation effects remains untested since LEP times. Future high-luminosity $\epem$ machines operating at the Z pole, such as the FCC-ee~\cite{FCCee} with $10^5$ times more data collected than at LEP, will feature negligible statistical uncertainties, and the latter QCD corrections will dominate the systematics of the $\AFBb$ measurement. Our understanding of QCD effects has significantly improved since the work of~\cite{Abbaneo:1998xt} thanks to theoretical and experimental progress that has been incorporated into modern PS$\,+\,$hadronization models. The main purpose of this work is to reanalyze the original measurements with modern MC parton-shower tools to see if the existing data-theory $\AFBb$ discrepancy could be (partly) reduced by a reevaluation of the impact of QCD effects.

The paper is organized as follows. In Section~\ref{sec:exp}, we succinctly review the LEP $\AFBb$ measurements and their uncertainties. In Section~\ref{sec:MC}, we describe the various modern parton shower MC event generators used here to simulate the LEP measurements and to reassess their associated QCD uncertainties. In Section~\ref{sec:results}, the eight LEP measurements are updated with the reevaluated QCD effects, a new  $\AFBb$ average value is derived, and perspectives at a future $\epem$ collider are outlined. In Section~\ref{sec:summary}, we conclude with a summary.

%%%%%%%%%%%%%%%%%%%%%%%%%%%%%%%%%%%%%%%%%%%%%%%%%%%%%%%%%%%%%%%%%%%%%%%%%%%%%%%%%%%%%%%%%%
%\section{Measurements and uncertainties of the $e^+e^- \to\bbbar$ FB asymmetry} %LEP Measurements vs. SM Predictions}
\section{Experimental measurements and uncertainties}
\label{sec:exp}

Eight measurements of $\AFBb$ were performed in $\epem$ collisions at LEP around the Z pole. All of them start by clustering the $\epem$ final-state particles into jets, mostly with the JADE algorithm~\cite{jade} and jet resolution parameter $y_{\rm cut}=0.01$ or 0.02, and  by identifying its quark flavour ($b$ tagging) to
%, and apply cuts on the jet invariant mass $\rm M_{jet}$ and/or on the jet resolution parameter $y_{\rm cut}=m^2_{ij}/E^2_{vis}$. 
determine the thrust axis of the event, used as a proxy of the $b$-quark direction. 
Secondly, the $b$ and $\bar{b}$ are separated through charge-tagging methods that rely on either lepton- or jet-charge measurements. On the one hand, in four measurements (ALEPH at $\sqrts = 91.21$~GeV~\cite{leptonALEPH}, DELPHI at $\sqrts = 91.26,91.23$~GeV~\cite{leptonDELPHI95}, L3 at $\sqrts = 91.26$~GeV~\cite{leptonL392}, and OPAL at $\sqrts = 91.25, 91.24$~GeV~\cite{leptonOPAL}), the original $b$ ($\bar{b}$) quark is identified from the negative (positive) charge of the leading lepton $\ell^\pm$ inside each jet (\ie\ that with the largest momentum perpendicular to the jet axis, through the $b\to B, b\to c\to D$ fragmentation(s), and subsequent $B,D\to\ell^\pm$ semileptonic decay). The raw FB asymmetry $\AFBobs$ is then obtained by fitting the corresponding distribution of polar angles $\theta$ between the $e^-$ and the thrust axis to the following expression
\begin{equation}
\frac{\mathrm{d}\sigma}{\mathrm{d}\cos\theta}=\sigma^{\rm tot}_{_{\bbbar}}\left(\frac{3}{8}(1+\cos^2\theta)+\AFBobs(1-2\chi_B)\cos\theta\right)\,,
\label{eq:fit_cos}
\end{equation}
%A final correction is applied $\AFBb=\AFBobs/(1-2\chi_B)$ 
where $\chi_B\approx 0.125$ is the $B^0\overline{B^0}$ effective mixing parameter, measured by each experiment {\it in situ}, encoding the probability that a semileptonically decaying $b$ quark is reconstructed as a $\overline{b}$~quark. 
In the other four measurements (ALEPH at $\sqrts = 91.23$~GeV~\cite{jetqALEPH}, DELPHI at $\sqrts = 91.23$~GeV~\cite{jetqDELPHI}, L3 at $\sqrts = 91.24$~GeV~\cite{jetqL3}, and OPAL at $\sqrts = 91.26$~GeV~\cite{jetqOPAL}), the $b$-quark charge is reconstructed from the constituent charged particles of the jet via $Q_{\rm jet}=\sum_i p_{\mathrm{L},i}^\kappa\,Q_i/\sum p_{\mathrm{L},i}^\kappa$ (where $p_{\mathrm{L},i}$ is the longitudinal momentum of the final-state particles, with individual charge $Q_i$, with respect to the thrust axis, and the power $\kappa$ varies between 0.4 and 0.6 depending on the experiment).
The $\AFBobs$ asymmetry is then extracted by fitting the $\cos\theta$ distribution to the expression
\begin{equation}
\frac{\mean{Q_F-Q_B}}{\mean{Q_{b}-Q_{\overline{b}}}} = \frac{8}{3}\AFBobs \frac{\cos\theta}{1+\cos^2\theta}\,,
\label{eq:fit_Q}
\end{equation}
where $Q_F$ ($Q_B$) is the jet charge of the forward (backward) hemisphere, $Q_b$ ($Q_{\overline{b}}$) is the jet charge in the hemisphere containing the $b$ ($\bar{b}$) quark. 

The measured $\AFBobs$ values are further corrected for different QCD effects, including higher-order perturbative QCD effects (hard gluon real radiation and virtual exchanges), 
as well as for further smearing due to $b$ and $(b\to)c$ soft and/or collinear radiation and hadronization, that lead to a difference between the result obtained with the reconstructed thrust axis (T) and the $b$-quark direction. This is done through the correction factor
\begin{equation}
\AFBbb=\AFBobs/(1-C_{\rm QCD})\,,\; {\rm with }\;\; C_{\rm QCD} = s_b\cdot C^{\rm T}_{\rm QCD}\,,
\label{eq:QCDcorr}
\end{equation}
where the full QCD effects are decomposed into the product of two coefficients: (i) the $C^{\rm T}_{\rm QCD}$ term including perturbative gluon radiation plus the non-perturbative conversion of partons into hadrons, determined originally at next-to-leading order (NLO) accuracy including hadronization effects, $C^{\rm T,had}_{\rm QCD} = (3.54\pm0.63)\%$~\cite{ALEPH:2005ab,Abbaneo:1998xt}, and known today at the partonic level at NNLO~\cite{Catani:1999nf} including $b$-quark mass effects: $C^{\rm T,parton}_{\rm QCD} = (3.92\pm0.25)\%$~\cite{Bernreuther:2016ccf}, and (ii) an extra experiment-dependent weighting factor,  $s_b\approx  0.3, 0.6$ for the jet- and lepton- charge-based extractions respectively, that accounts for the convolution of the detector response with any biases introduced in the event topology by the data selection and analysis method of each experiment. In practice, the overall QCD correction $C_{\rm QCD}$ and its uncertainty,
%accounting for both hard gluon radiation and non-perturbative fragmentation effects, 
were determined from the spread of the differences between the parton- and hadron-level results obtained with various tunes of the \jetset~7.408~\cite{pythia} parton shower simulations~\cite{Abbaneo:1998xt}, after confirming that the latter (at the parton level) were consistent with the $C^{\rm T,parton}_{\rm QCD}$ value obtained analytically at NLO. %The main motivation of this work is to revise those estimates. 
Lastly, the fitted asymmetries require a final correction for Quantum Electrodynamics (QED) and electroweak effects, including $\gamma$ radiation and exchange, Z-$\gamma$ interference, and a shift to the pole $m_{\rm Z} = 91.1880$~GeV mass. Accounting for the latter effects increases the observed asymmetry by 1.9\% with negligible uncertainties~\cite{zfitter}.
\begin{equation}
\AFBb = \AFBbb + \delta\AFBbb,\;\; {\rm with }\;\; \delta\AFBbb = +1.9\%.
\label{eq:QEDcorr}
\end{equation}

Table~\ref{tab:AFBb} lists the eight individual $\AFBb$ measurements with the breakdown of their uncertainties.
The statistical uncertainties dominate, being about twice bigger than the systematic ones, while the QCD uncertainties, fully correlated among all measurements, account for about half of the latter. The quoted QCD uncertainties are not always exactly (although they are close to) those quoted in each original LEP paper, but instead include in quadrature all sources of QCD-related effects extracted from the original measurements in the final $\AFBb$ averaging study~\cite{ALEPH:2005ab}.
The combination of the eight measurements gives $\AFBb = 0.0992 \pm 0.0016$, %(with $\pm1.6\%$ global uncertainty) 
with 1.5\% statistical uncertainty, and 0.7\% systematic uncertainty dominated by $\sim$0.5\% QCD sources (taken as fully correlated in the final average).

\begin{table}[htbp!]
\caption[]{LEP measurements of $\AFBb$ and associated relative statistical, QCD-related and total systematic uncertainties. The last row lists the LEP combined result.
\label{tab:AFBb}}
\centering
\resizebox{0.95\textwidth}{!}{
\begin{tabular}{lcccc}\hline
Measurement:  &  $(\AFBb) \pm \delta\mathrm{(stat)} \pm \delta\mathrm{(syst)}$ & \multicolumn{3}{c}{relative uncertainties}\\
\;\;Experiment &    &  stat. & QCD syst. & total syst. \\\hline
Lepton-charge based: & & & & \\
\;\;ALEPH (2002)~\cite{ALEPH:2005ab,leptonALEPH} & $0.1003 \pm 0.0038 \pm 0.0017$ &  $3.8\%$  & $0.7\%$ & $1.7\%$\\
\;\;DELPHI  (2004--05)~\cite{ALEPH:2005ab,leptonDELPHI95} & $0.1025 \pm 0.0051 \pm 0.0024$ & $5.0\%$  & $1.2\%$ & $2.3\%$\\ 
\;\;L3  (1992--99)~\cite{ALEPH:2005ab,leptonL392} & $0.1001 \pm 0.0060 \pm 0.0035$ & $6.0\%$ & $1.8\%$ & $3.5\%$ \\
\;\;OPAL  (2003)~\cite{ALEPH:2005ab,leptonOPAL}& $0.0977 \pm 0.0038 \pm 0.0018$ & $3.9\%$ & $1.1\%$ & $1.8\%$ \\\hline
Jet-charge based: & & & & \\
\;\;ALEPH (2001)~\cite{ALEPH:2005ab,jetqALEPH} & $0.1010 \pm 0.0025 \pm 0.0012$ & $2.5\%$ & $0.7\%$ & $1.2\%$ \\
\;\;DELPHI (2005)~\cite{ALEPH:2005ab,jetqDELPHI} & $0.0978 \pm 0.0030 \pm 0.0015$ & $3.1\%$ & $0.7\%$& $1.5\%$\\
\;\;L3  (1998)~\cite{ALEPH:2005ab,jetqL3} & $0.0948 \pm 0.0101 \pm 0.0056$ & $10.6\%$ & $4.3\%$ & $5.9\%$ \\ 
\;\;OPAL (1997,2002)~\cite{ALEPH:2005ab,jetqOPAL}& $0.0994 \pm 0.0034 \pm 0.0018 $& $3.4\%$ & $0.7\%$ & $1.8\%$\\\hline
Combination~\cite{ALEPH:2005ab}  & $0.0992 \pm 0.0015 \pm 0.0007$ & 1.5\% & 0.5\%  & 0.7\% \\\hline
\end{tabular}
}
\end{table}

%%%%%%%%%%%%%%%%%%%%%%%%%%%%%%%%%%%%%%%%%%%%%%%%%%%%%%%%%%%%%%%%%%%%%%%%%%%%%%%%%%%%%%%%%%
\section{Monte Carlo simulations}
\label{sec:MC}

In order to replicate in simulation each one of the eight $\AFBbb$ extractions at LEP, and reassess the size of the QCD systematic uncertainties with more modern MC tools, we use the \pythia\,8.226~\cite{pythia8} event generator, with seven different sets of $\epem$ parameters (``tunes'') to model the showering and hadronization, as well as the \pythia\,8.210 MC generator combined with two versions of the alternative \vincia\ parton shower~\cite{vincia1,vincia2}. The QCD evolution in the \pythia\,8 MC generator is based on Dokshitzer--Gribov--Lipatov--Altarelli--Parisi (DGLAP) LO splittings~\cite{dglap} with the option to use the Catani--Marchesini--Webber (CMW) rescaling of the strong coupling $\alphas$~\cite{CMW} to account for NLO corrections to soft gluon emissions, approximating soft and collinear radiation beyond the leading logarithm in an effective way. The $b$ and $c$ quark fragmentation is modeled according to the Lund string model~\cite{Andersson:1983ia} combined with the Bowler~\cite{Bowler:1981sb} function or, at LEP-1 times, to the Peterson function~\cite{Peterson:1982ak}.
The \vincia\ parton shower is based on the dipole, or antenna, picture of QCD radiation~\cite{Gustafson:1987rq} that follows a $2 \to 3$ branching evolution (\eg\ $\qqbar\to q g \bar{q}$), rather than the standard $1\to2$ splitting functions typical of collinear factorization (\pythia\,8 combines a $1\to 2$ splitting probability with a $2\to3$ phase-space mapping). By considering colour dipoles, \ie\ by considering particle emissions as stemming
from colour-anticolour pairs, the \vincia\ shower effectively incorporates coherent (wide-angle) emissions in the DGLAP formalism to leading power in $(1/N_c^2)$, where $N_c$ is the number of colours. Compared to \pythia\,8 stand-alone, the \vincia\ dipole shower effectively %covering extended regions of phase space
includes suppressed unordered branchings that cover the hard region of phase space, as well as systematic ``next-to-leading-colour'' corrections~\cite{vincia1}. In the non-perturbative sector, \vincia\ uses the fragmentation model of \pythia\,8 but, given the aforementioned differences between the two parton evolutions that the hadronization model does not reabsorb automatically in the soft region, its parameters (in particular the $b$-quark fragmentation) are retuned. Such differences effectively lead to modified $b$-jet thrust axis reconstruction, and to variations in different regions of the (refitted) $b\to B$ fragmentation~\cite{mcplots}. The two \vincia~1.1~\cite{vincia1} and 2.2~\cite{vincia2} versions used, differ basically in the fact that the former is tuned to reproduce $\epem$ data alone, whereas the latter includes also the latest results from proton-proton collisions at the LHC.

\begin{table}[htbp!]
\caption{Details of the nine different sets of MC parameter settings in the \pythia\,8 stand-alone and \vincia\ parton-shower simulations used in this work.
\label{tab:tunes}}
\centering
\begin{tabular}{l p{13.6cm}}\hline
\pythia\,8.226: & \\%\hline
Tune-1* & First \pythia\,8 parameter set (2006), based on 1990s LEP-1 studies with {\sc jetset}~\cite{pythia}, modified (`*' mark) to use the Peterson fragmentation, rather than the default Lund-Bowler, function to approximate the original LEP QCD studies~\cite{Abbaneo:1998xt}.\\%\hline
Tune-2 & 2007 tune to LEP-1 particle composition.  Bowler $b$-frag.\,parameter: 0.58\\%\hline
Tune-3 & 2009 tune of hadronization and timelike-shower parameters using a large set of LEP-1 data with {\sc Rivet$\,+\,$Professor}~\cite{rivet}. Bowler $b$-frag.\,parameter: 0.8 \\%\hline
Tune-4 & 2013 tune of shower and hadronization parameters with LEP data. CMW~$\alphas$~scale.\\%\hline
Tune-5 & 2013 flavour-composition tune based on ALEPH$\,+\,$DELPHI event shapes, ALEPH identified hadron spectra, PDG multiplicities, and ALEPH $b$ fragmentation.\\%\hline
Tune-6 & Second update of Tune-5.\\%\hline
Tune-7 & 2013 Monash tune~\cite{monash} using $\epem$, pp, p$\rm \bar{p}$ data. Bowler $b$-frag.\,parameter: 0.855\\\hline
\vincia: & \\%\hline
1.1 & 2013 default tune to LEP-1 event shapes, jet rates, multiplicities, inclusive spectrum. 1-loop CMW $\alphas$ evolution.
Bowler $b$-fragmentation parameter: 0.9.\\%\hline
2.2 & 2016 update of 1.1 tunes. 2-loop CMW $\alphas$ running. Bowler $b$-frag.\,parameter: 1.1\\\hline
\end{tabular}
\end{table}

Details of the seven \pythia\,8 and the two \vincia\ tunes used are listed in Table~\ref{tab:tunes}. Most of them, except \pythia\,8 Tune-7 and \vincia~2.2 that also use hadron-collider data, employ exclusively $\epem$ results 
to constrain the parameters of the underlying showering and hadronization models. The closest parameter set to that used in the original LEP studies of the QCD corrections and uncertainties of the FB asymmetry~\cite{Abbaneo:1998xt} is \pythia\,8 Tune-1*, where the `*' superscript indicates that we have modified the heavy-quark fragmentation from the default Lund-Bowler to the Peterson function to approximate the LEP-1 settings\footnote{We checked the results of varying the parameter of the Peterson function over $\epsilon_b=0.005\pm0.003$ to cover the range of the JETSET 7.4 settings as well as of more recent fits of the LEP-1 $b$-jet data~\cite{DELPHI:2011aa}, and we obtain final $\AFBobs$ values for the Tune-1* lepton-charge analyses that vary within $\pm1\%$ (they are plotted as a larger vertical bar on the first point of each canvas of Fig.~\ref{fig:lepton_AFBobs_vs_MC}).}. The whole MC setup effectively corresponds to nine different modelings of the underlying QCD effects: parton radiation, and heavy-quark, light-quark and gluon hadronization.

\begin{figure}[htbp!]
\centering
\includegraphics[width=0.49\columnwidth]{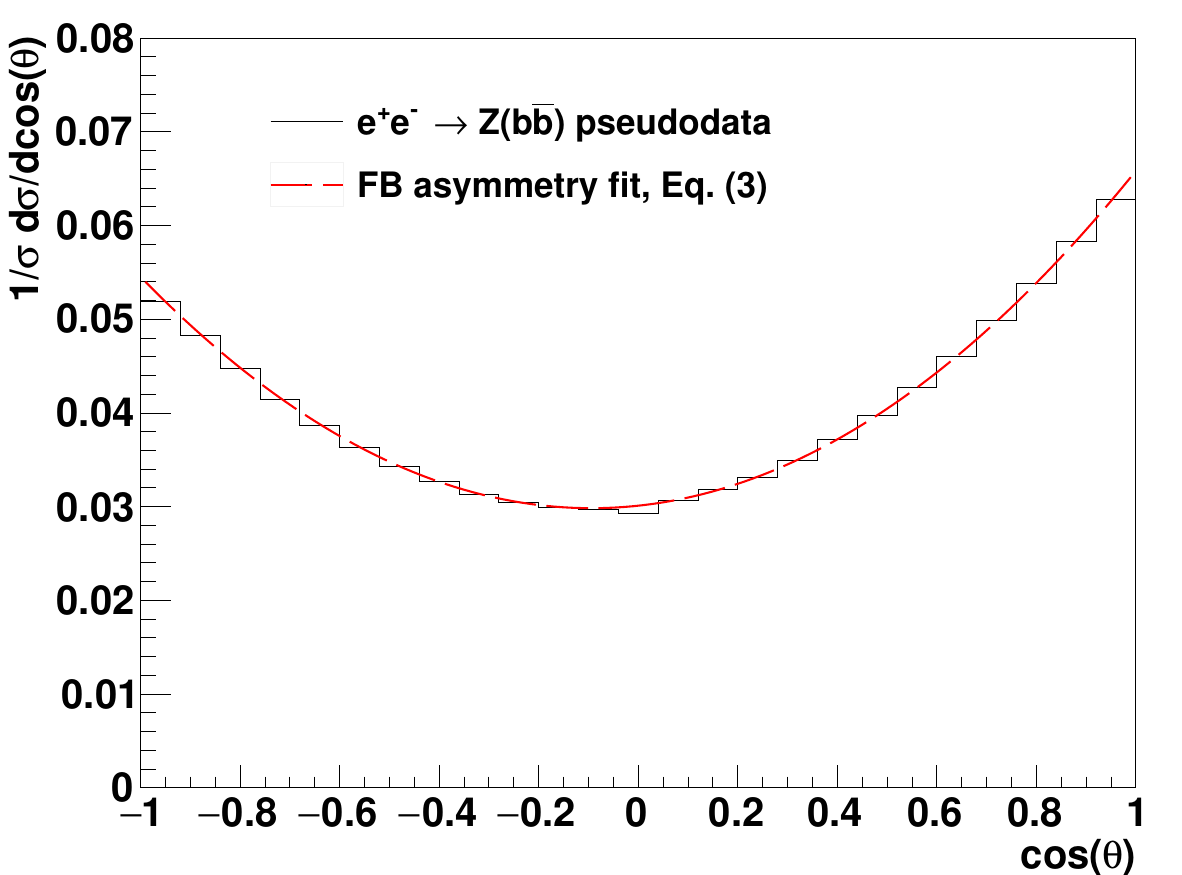}
\includegraphics[width=0.49\columnwidth]{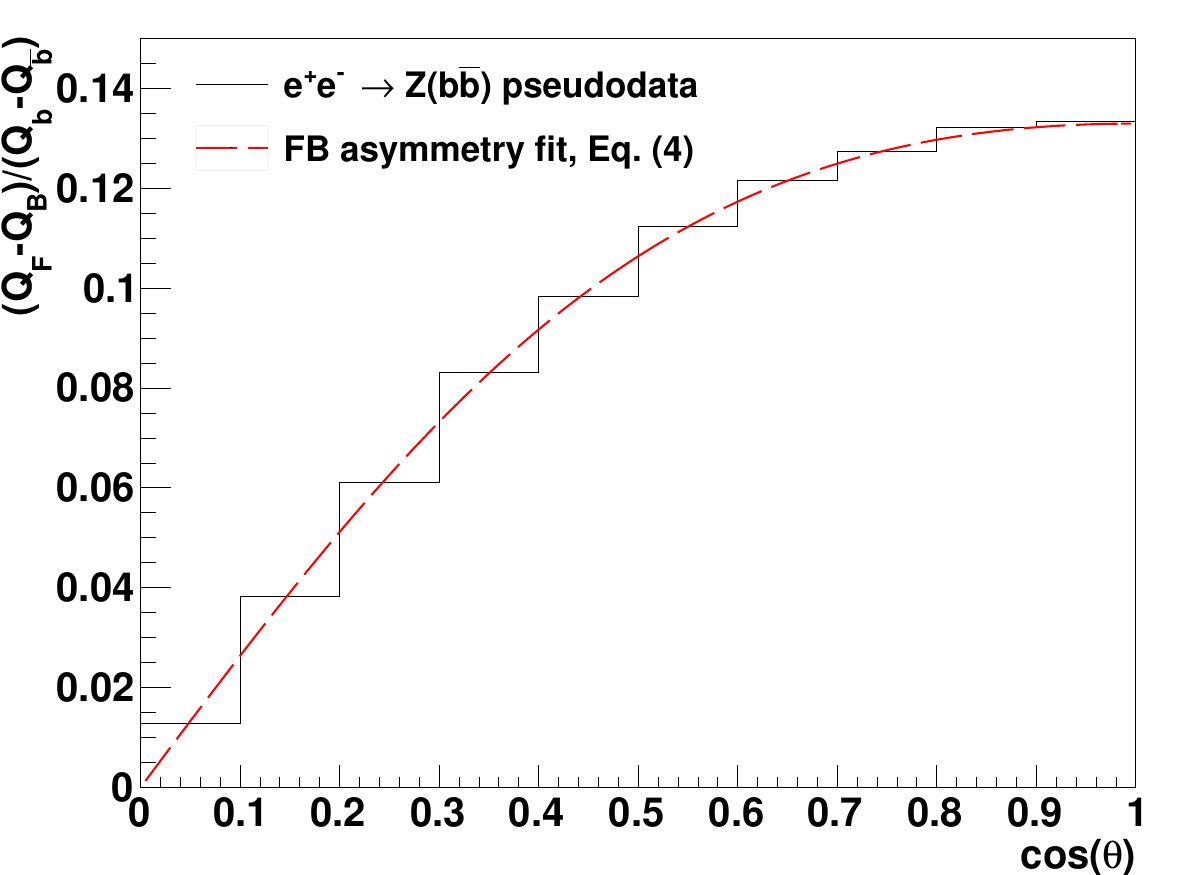}
\caption{Examples of simulated $\cos\theta$ distributions in $\epem\to \mathrm{Z}(\bbbar)$ events generated with \pythia~8 (tune 1$^*$). Left: Asymmetry reconstructed from the $b$-quark lepton charge fitted to Eq.~(\protect\ref{eq:fit_cos}) for a L3-like analysis. Right: Asymmetry reconstructed from the $b$-quark charge, for events passing the OPAL analysis flow, fitted to Eq.~(\protect\ref{eq:fit_Q}).
\label{fig:fit_examples}}
\end{figure}

The eight original LEP analyses of $\epem\to \mathrm{Z}(\bbbar)$ data and associated $\AFBb$ extractions have been implemented in a MC event simulation based on \pythia\,8.226, stand-alone or with \vincia, with the nine different tunes listed in Table~\ref{tab:tunes}. One-hundred million $\epem\to \mathrm{Z}(\bbbar)$ events are generated at $\sqrts = 91.24$~GeV with QED radiation switched on, in order to mimic closely the original experimental conditions, and analyzed as done in the original experiments by implementing as close as possible the analysis flow and selection criteria of the lepton- and jet-charge analyses of each one of the eight measurements\footnote{Obviously, our analysis cannot achieve the results of the full simulations of the detector-dependent responses of the original studies and, in particular, the associated $s_b$ weights of the QCD correction factors in Eq.~(\ref{eq:QCDcorr}) cannot be directly determined.}. First, all final-state particles (except neutrinos) are clustered into jets %\footnote{We note that the actual experimental analyses dealt originally with a mixture of various quark  %flavours (Z\,$\to\uubar,\ddbar,\cdots$) in the final state, flavours (Z\,$\to\qqbar$) in the final state, whereas our simulations start off with a pure $\bbbar$ sample, and therefore any potential experimental effect on the $b$-tagging efficiency and/or misidentification rates that was left originally uncorrected for is not covered by our estimates.} 
using the \jade\ algorithm, interfaced via \fastjet~\cite{fastjet}, with resolution parameter $y_{\text{cut}} = m_{\rm jet}^2/E_{\rm vis}^2= 0.01$ (DELPHI) or 0.02 (rest of experiments). The standard thrust axis of the event is then computed employing all final-state particles except neutrinos\footnote{The change in the QCD correction to $\AFBbb$ due to undetected neutrinos is only of about 0.002~\cite{Abbaneo:1998xt}.}, as a proxy of the original $b$-quark direction. The same original jet invariant mass $m_{\rm jet}> 6$~GeV selection criteria (ALEPH and L3), as well as the $\kappa = 0.4$ (L3), 0.5 (OPAL and ALEPH), 0.6 (DELPHI) exponents (for the jet-charge extraction), and the (transverse) momenta ($p_T$) $p$ cuts (above 2, 2.5, or 4 GeV) on the final electron and muons (for the lepton-based analyses), are applied. The resulting charged-lepton and jet-charge angular distributions, with the same $\cos\theta$-binning as used in the original measurements, are then fitted to Eqs.~(\ref{eq:fit_cos}) and (\ref{eq:fit_Q}) respectively. Examples of the fitted $\cos\theta$ distributions are shown in Fig.~\ref{fig:fit_examples} for lepton- (left) and jet- (right) charge analyses with $\epem\to \mathrm{Z}(\bbbar)$ events generated with \pythia~8 (tune 1$^*$). In the lepton-based analyses, the extracted $\AFBobs$ fit values are corrected by the specific $\chi_B=0.1177$--0.1318 $B$-mixing factors directly measured by each experiment (the ALEPH and OPAL $\chi_B$ values are updated by half-a-percent, following the final corrections discussed in~\cite{ALEPH:2005ab}).

%%%%%%%%%%%%%%%%%%%%%%%%%%%%%%%%%%%%%%%%%%%%%%%%%%%%%%%%%%%%%%%%%%%%%%%%%%%%%%%%%%%%%%%%%%
\section{Results and discussion}
\label{sec:results}

\begin{figure}[htpb!]
\centering
\includegraphics[width=0.47\linewidth,height=4.8cm]{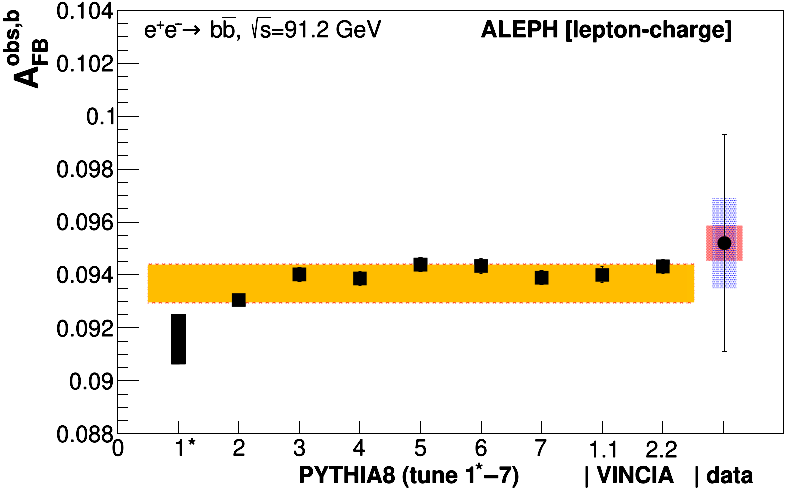}
\includegraphics[width=0.47\linewidth,height=4.8cm]{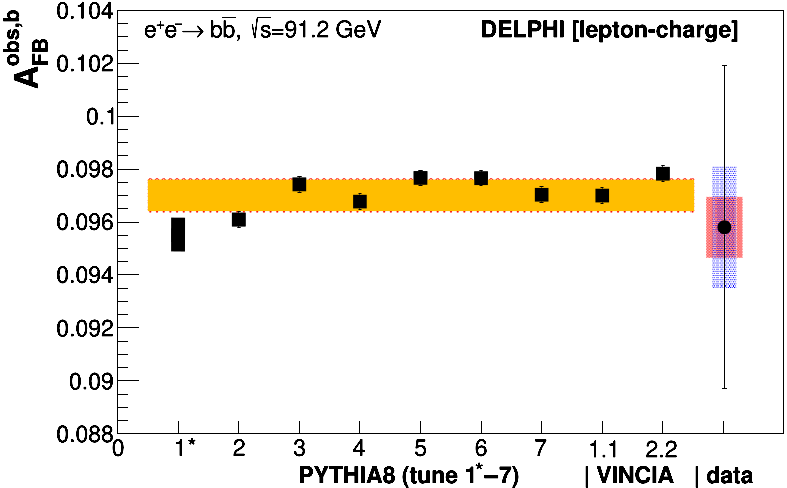}\\
\includegraphics[width=0.47\linewidth,height=4.8cm]{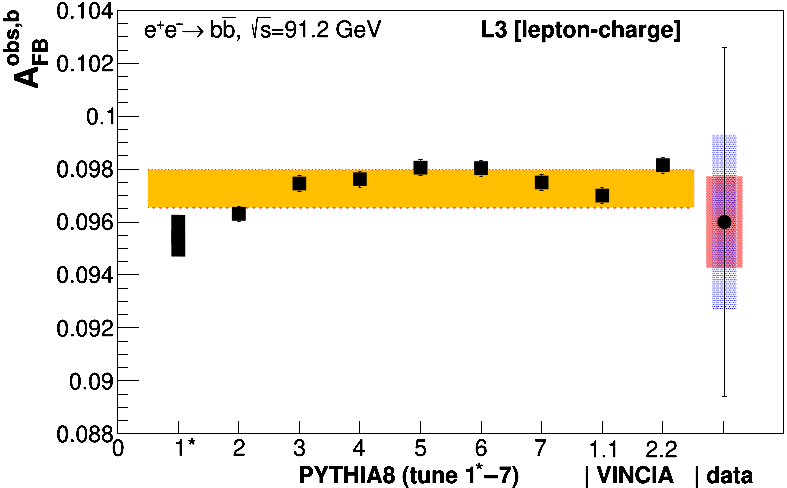}
\includegraphics[width=0.47\linewidth,height=4.8cm]{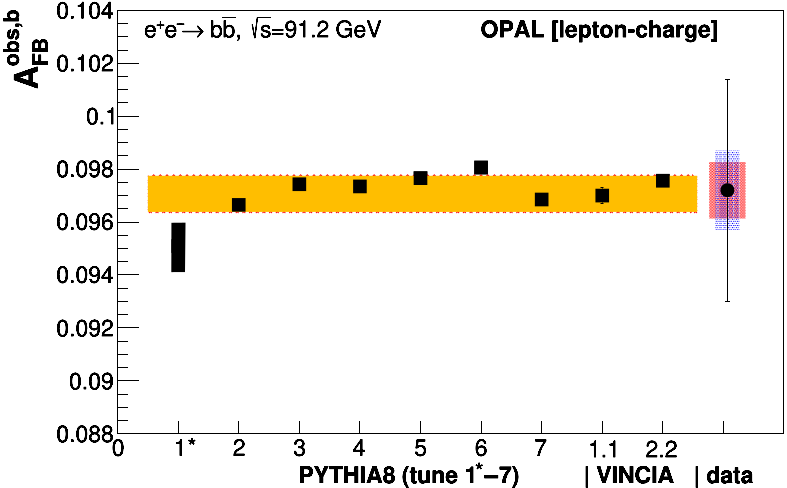}
\caption[]{Values of the $b$-quark forward-backward asymmetry extracted from lepton-charge analyses of $\epem\to \mathrm{Z}(\bbbar)$ simulations (black rectangles) based on seven \pythia\,8 and two \vincia\ tunes, compared to the corresponding experimental results (rightmost black circle point with statistical uncertainties indicated by the error bar, and QCD, in magenta, and total, in blue, systematic uncertainty boxes) measured by ALEPH (top left)~\cite{leptonALEPH}, DELPHI (top right)~\cite{leptonDELPHI95}, L3 (bottom left)~\cite{leptonL392}, and OPAL (bottom right)~\cite{leptonOPAL}. The orange band around the MC points indicates their overall assigned uncertainty and the outer hatched red band includes the NNLO pQCD uncertainty ($\pm0.25\%$) added in quadrature, as explained in the text.}
\label{fig:lepton_AFBobs_vs_MC}
\end{figure}

\begin{figure}[htpb!]
\centering
\includegraphics[width=0.47\linewidth,height=4.8cm]{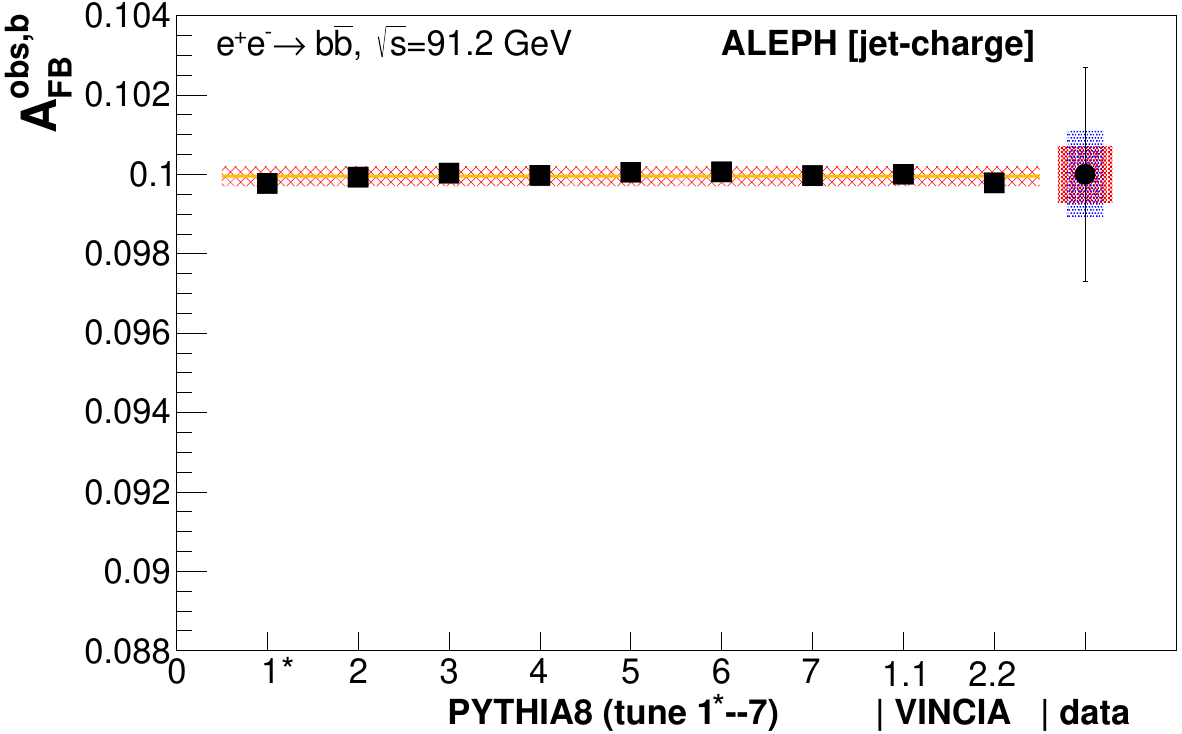}
\includegraphics[width=0.47\linewidth,height=4.8cm]{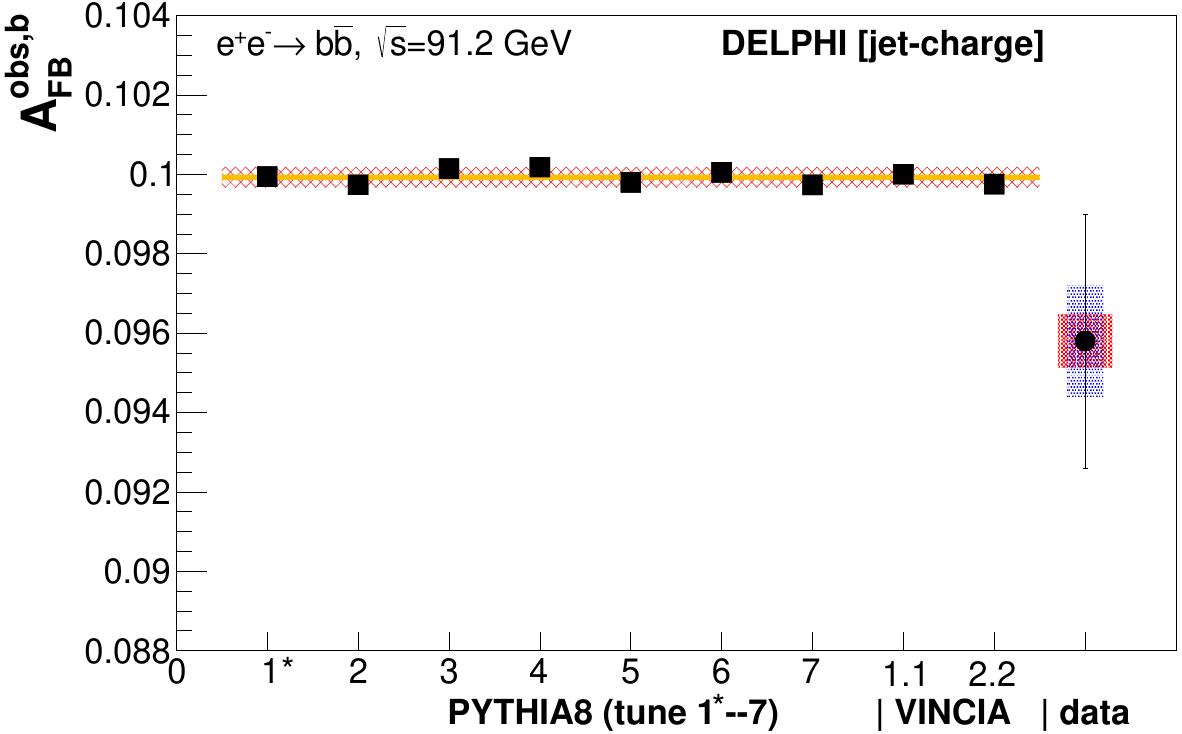}\\
\includegraphics[width=0.47\linewidth,height=4.8cm]{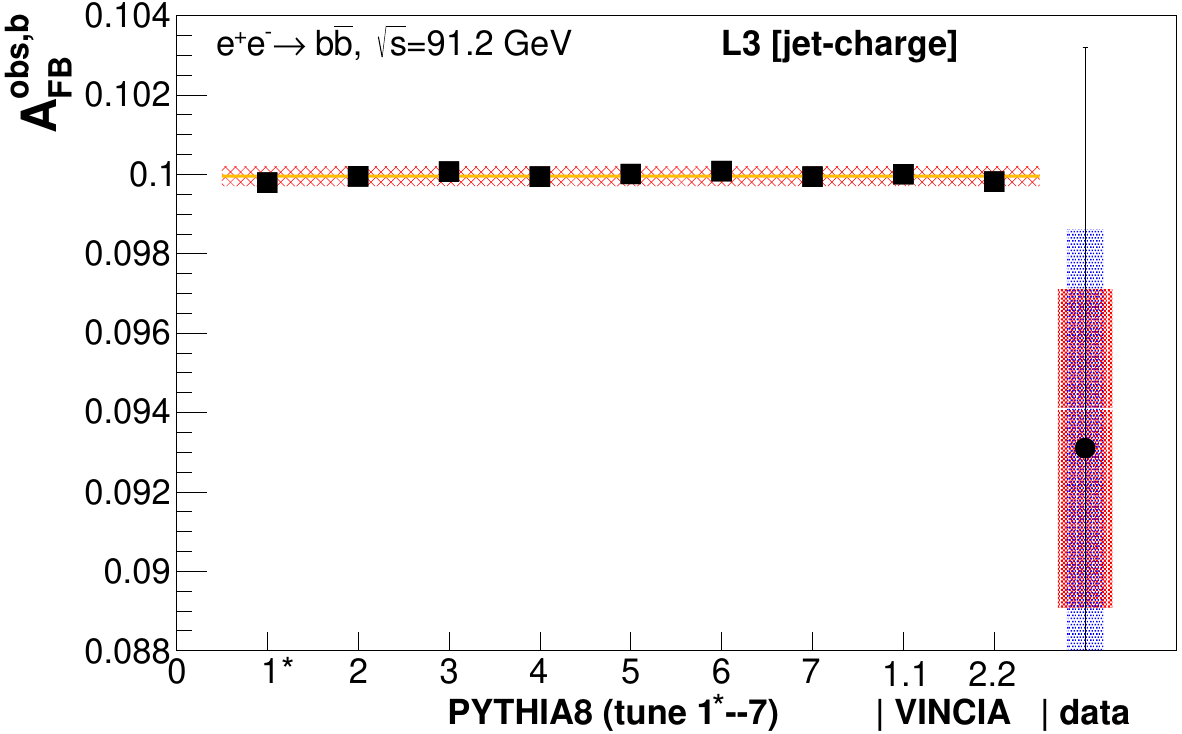}
\includegraphics[width=0.47\linewidth,height=4.8cm]{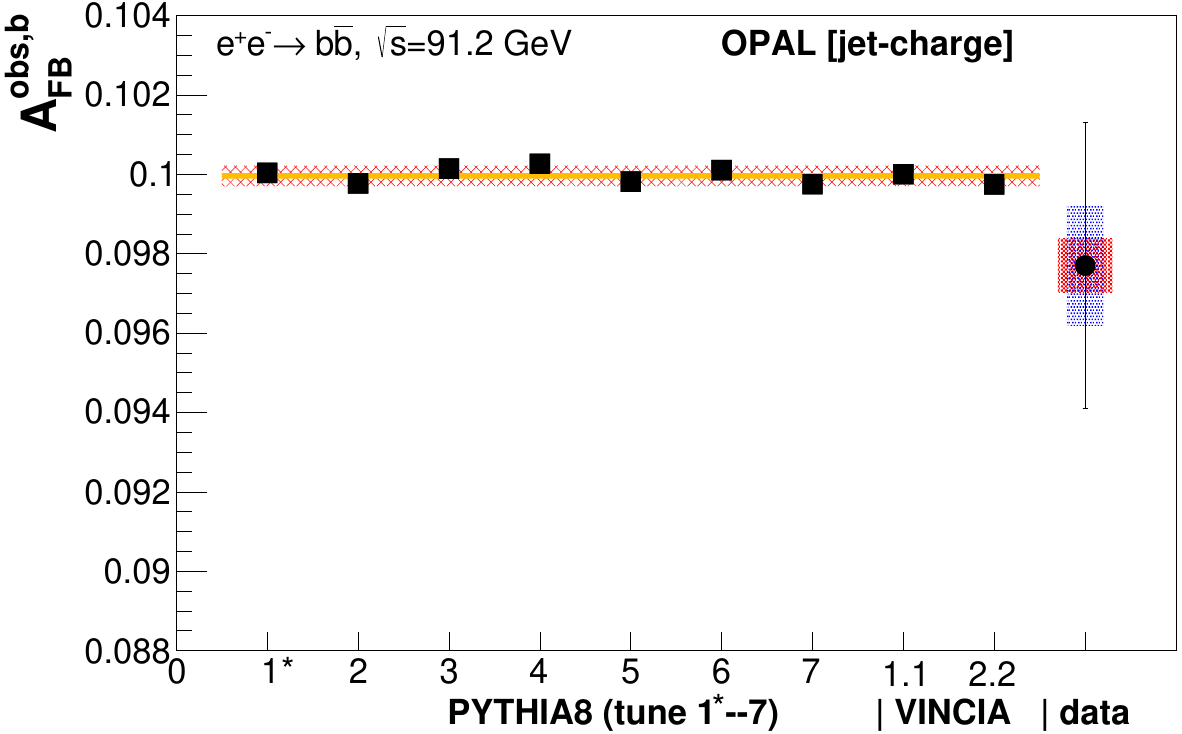}
\caption[]{Values of the $b$-quark forward-backward asymmetry extracted from jet-charge analyses of $\epem\to \mathrm{Z}(\bbbar)$ simulations (black squares) based on seven \pythia\,8 and two \vincia\ tunes, compared to the corresponding experimental results (rightmost black circle point with statistical uncertainties indicated by the error bar, and QCD, in magenta, and total, in blue, systematic uncertainty boxes) measured by ALEPH (top left)~\cite{jetqALEPH}, DELPHI (top right)~\cite{jetqDELPHI}, L3 (bottom left)~\cite{jetqL3},and OPAL (bottom right)~\cite{jetqOPAL}. The orange band around the MC points indicates their overall assigned uncertainty and the outer hatched red band includes the NNLO pQCD uncertainty ($\pm0.25\%$) added in quadrature, as explained in the text.} 
\label{fig:jet_AFBobs_vs_MC}
\end{figure}

Through the procedure described above, nine different MC values of $\AFBobs$ are extracted for each one of the eight experimental setups simulated, which we compare among themselves as well as against the experimental data in Fig.~\ref{fig:lepton_AFBobs_vs_MC} (lepton-based extractions) and Fig.~\ref{fig:jet_AFBobs_vs_MC} (jet-charge based analyses). We recall, first, that the experimental $\AFBobs$ values plotted differ slightly from the corresponding $\AFBb$ values quoted in Table~\ref{tab:AFBb} since, as aforementioned, the latter are further corrected for electroweak effects and shifted to the pole $m_{\rm Z}$ mass. Since we are interested in comparing the uncertainties in the data and MC simulation, with the latter estimated from the {\it relative} differences among the $\AFBobs$ values extracted using different PS$\,+\,$hadronization models, we do not correct for the latter effects in these plots. %\footnote{We note, in any case, that if one would apply the correction given by  Eq.~(\ref{eq:QEDcorr}) to our MC results, the obtained $\AFBb$ values would all approach the theoretical $(\AFBb)_{_{\rm th}} = 0.1030 \pm 0.0002$ prediction.}. 
The large MC data samples generated (100 million events, of which between 20--30 million events pass the analysis criteria given by the concrete experiment-dependent angular and momenta cuts applied on every final-state particle) lead to very small statistical fluctuations in the simulated asymmetries, and any relative differences among them is in principle just related to intrinsic differences in the physics modelling and settings of each tune. 

The FB asymmetries plotted in Figs.~\ref{fig:lepton_AFBobs_vs_MC} and \ref{fig:jet_AFBobs_vs_MC} show that, within the (comparatively larger) experimental uncertainties, data and simulation results agree well for all systems. %The first (leftmost) MC point corresponds to the \pythia\,8 Tune-1* result obtained with the 1990 \jetset\ parameter set, which basically corresponds to the baseline MC {\sc jetset}~7.408 tune employed in the determination of the original QCD parton-shower and hadronization uncertainties at LEP~\cite{Abbaneo:1998xt}. 
The overall LEP QCD uncertainty of $\AFBobs$ was derived from the average of the differences between the QCD corrections at hadron- and parton-level based on the results of the baseline MC {\sc jetset}~7.408 tune (similar to the leftmost \pythia~8 tune-1* MC point in both figures), four other %experiment-dependent 
{\sc jetset} variations, and one \herwig~5.1~\cite{herwig} simulation. Here, we similarly estimate the PS$\,+\,$hadronization uncertainties of the bottom-quark asymmetry from the spread of the $\AFBobs$ results obtained from the nine MC tunes used.
%The spread of the different parton-shower and hadronization models is,
This is shown as an orange band around the MC points, which corresponds to the 95\% (68\%) confidence-level standard deviation of the predictions for lepton- (jet-) charge analysis, which we take as indicative of the associated overall uncertainty for each system. For the lepton-based analyses, we take a conservative $2\sigma$ error, rather than the standard $1\sigma$ used for the jet-charge case, to cover the comparatively lower asymmetry value (and larger associated uncertainties) derived for the tune-1*. The so-derived $\AFBobs$ uncertainties amount to $\sim$0.7\% and $\sim$0.05\% for the lepton- and jet- charge measurements. The fact that the results of the jet-based analyses have much smaller spread than those from the lepton-based ones is not unexpected: Although, in principle, leptonic final states allow for a better $b$-quark tagging and more accurate charge-sign determination than the pure hadronic final-state analyses, they are also prone to larger QCD-related systematic uncertainties because of a larger sensitivity to the $b$- and $c$-quark fragmentation details compared to the jet-charge extractions.
The LEP data points plotted in Figs.~\ref{fig:lepton_AFBobs_vs_MC} and \ref{fig:jet_AFBobs_vs_MC} show full QCD uncertainties (magenta boxes) that include also a 0.63\% contribution from the NLO perturbative $C^{\rm T,had}_{\rm QCD}$ correction factor given by Eq.~(\ref{eq:QCDcorr}). In order to make a more accurate comparison of the original and the new QCD uncertainties, we add quadratically to the parton-shower orange band the $\pm0.25\%$ partonic NNLO uncertainty~\cite{Bernreuther:2016ccf}. Our final reevaluated full QCD uncertainty is shown with the outer hatched red band around the MC points. The impact of this extra pQCD uncertainty is barely visible for the lepton-based analyses, dominated by the $\sim$0.7\% PS$\,+\,$hadronization uncertainties, but it becomes now the dominant source for the jet-based studies. The rightmost columns of Table~\ref{tab:AFBbnew} list the updated QCD (and, correspondingly, total) systematic uncertainties for each one of the LEP $\AFBbb$ extractions.

\begin{table}[htbp!]
\caption[]{Updated values of the $\AFBb$ asymmetries extracted at each LEP experiment (listed in Table~\ref{tab:AFBb}) with the central value slightly increased to account for NNLO pQCD corrections~\cite{Bernreuther:2016ccf}, and with the QCD-related (and, accordingly, total) systematic uncertainties updated in this work. The last row lists the average result derived combining the individual estimates as described in the text.
\label{tab:AFBbnew}}
\vspace{-0.35cm}
\centering
\resizebox{0.9\textwidth}{!}{
\begin{tabular}{lcccc}\hline
Extraction:  &  $(\AFBb) \pm \delta\mathrm{(stat)} \pm \delta\mathrm{(syst)}$ & \multicolumn{3}{c}{relative uncertainties}\\
\;\;Experiment &    &  stat. & QCD syst. & total syst. \\\hline
Lepton-charge based: & & & & \\
\;\;ALEPH (2002)       & $0.1006 \pm 0.0038 \pm 0.0017$ & $3.8\%$ & $0.7\%$ & $1.7\%$\\
\;\;DELPHI  (2004--05) & $0.1027 \pm 0.0051 \pm 0.0021$ & $5.0\%$ & $0.7\%$ & $2.0\%$\\ 
\;\;L3  (1992--99)     & $0.1003 \pm 0.0060 \pm 0.0030$ & $6.0\%$ & $0.7\%$ & $3.0\%$ \\
\;\;OPAL  (2003)       & $0.0980 \pm 0.0038 \pm 0.0014$ & $3.9\%$ & $0.7\%$ & $1.6\%$ \\\hline
Jet-charge based: & & & & \\
\;\;ALEPH (2001)     & $0.1011 \pm 0.0025 \pm 0.0010$ &  $2.5\%$ & $0.3\%$ & $1.0\%$ \\
\;\;DELPHI (2005)    & $0.0979 \pm 0.0030 \pm 0.0013$ &  $3.1\%$ & $0.3\%$ & $1.3\%$\\
\;\;L3  (1998)       & $0.0949 \pm 0.0101 \pm 0.0039$ & $10.6\%$ & $0.3\%$ & $4.0\%$ \\ 
\;\;OPAL (1997,2002) & $0.0995 \pm 0.0034 \pm 0.0017$ &  $3.4\%$ & $0.3\%$ & $1.7\%$ \\\hline

Combination          & $0.0995 \pm 0.0015 \pm 0.0006$ &  $1.5\%$ & $0.3\%$ & $0.6\%$ \\\hline
\end{tabular}
}
\end{table}
%... Blue->PrintEst:                Value     Full   Stat    Syst uncorr.    QCD corr.  
%... Blue->PrintEst( 0) ALEPH(lep):  0.1006 +- 0.0042 (0.0038 +- 0.0016 +- 0.0008)
%... Blue->PrintEst( 1) DELPHI(lep):  0.1027 +- 0.0056 (0.0051 +- 0.0021 +- 0.0007)
%... Blue->PrintEst( 2) L3(lep):  0.1003 +- 0.0068 (0.0060 +- 0.0030 +- 0.0008)
%... Blue->PrintEst( 3) OPAL(lep):  0.0980 +- 0.0042 (0.0038 +- 0.0014 +- 0.0008)
%... Blue->PrintEst( 4) ALEPH(jet):  0.1011 +- 0.0027 (0.0025 +- 0.0010 +- 0.0003)
%... Blue->PrintEst( 5) DELPHI(jet):  0.0979 +- 0.0033 (0.0030 +- 0.0013 +- 0.0003)
%... Blue->PrintEst( 6) L3(jet):  0.0949 +- 0.0108 (0.0101 +- 0.0038 +- 0.0003)
%... Blue->PrintEst( 7) OPAL(jet):  0.0995 +- 0.0038 (0.0034 +- 0.0017 +- 0.0003)

As a last exercise, we can try to assess the impact that our revised QCD uncertainties, combined with the updated massive NNLO pQCD correction of~\cite{Bernreuther:2016ccf}, would have on the final LEP $\AFBb$ average. Of course, such an exercise is simplistic and cannot be compared to the full LEP $\AFBb$ combination~\cite{ALEPH:2005ab}, where a global fit of all heavy flavour measurements --- including $R_b$, charm observables, mixing, semileptonic branching ratios --- was performed. In addition, an updated realistic full simulation would likely lead to modifications of the detector-dependent $s_b$ bias factors of the NNLO pQCD correction in Eq.~(\ref{eq:QCDcorr}). Still, it is an informative procedure to estimate the numerical impact that the updated QCD effects considered here would naively have on the $\AFBb$ weighted average. If, in Eq.~(\ref{eq:QCDcorr}), one takes into account the complete massive NNLO correction (rather than the NLO one used at LEP times) weighted by the $s_b\approx 0.3, 0.6$ experimental bias factors determined in the original lepton- and jet- charge analyses, the central value of the each extracted FB asymmetry increases slightly by about 0.3\% and 0.1\% for lepton- and jet- charge analyses, respectively. If, in addition, we use the new QCD uncertainties estimated as explained above, of order $0.7\%\otimes0.25\%$ (for the lepton-charge analysis) and $0.05\%\otimes0.25\%$ (for the jet-charge extraction), we obtain the new individual $\AFBbb$ values listed in Table~\ref{tab:AFBbnew}. To obtain a single final asymmetry from all these results, we use the Best Linear Unbiased Estimate (BLUE) method as implemented in the {\sc blue} (v.2.1.1) code~\cite{blue}, combining all individual $\AFBb$ values with their (newly reassessed) QCD uncertainties taken as fully correlated among experiments, and their statistical and remaining systematics as fully uncorrelated. 
Such a procedure increases the combined $b$-quark FB asymmetry by $+0.0003$ (with a total uncertainty, dominated by statistical uncertainties, basically unmodified) compared to the average derived with the BLUE method applied on the original LEP asymmetries listed in Table~\ref{tab:AFBb}. The corresponding new result $\AFBb = 0.0995\pm0.0016$ would slightly diminish the tension between the $b$-quark asymmetry and the SM fit from 2.4$\sigma$ to 2.2$\sigma$. Such a reduction of the data-theory discrepancy would also propagate into the weak mixing angle derived from $\AFBb$, decreasing by an equal amount the existing pull between $\weakeffb$ and the current $\weakeff$ world average. 

The results presented here indicate that further improving the theoretical calculations and reducing the associated QCD systematic uncertainties alone will never solve this long-term discrepancy, because the experimentally measured $\AFBb$ precision is ultimately limited by a dominant $\pm$1.5\% statistical uncertainty at LEP. Eventual confirmation or resolution of this long-term discrepancy requires therefore a new high-luminosity $\epem$ collider collecting orders-of-magnitude more data at the Z pole. At the FCC-ee, with $6\cdot 10^{12}$ Z bosons expected (compared to $2\cdot 10^{7}$ Z's at LEP), the $\AFBb$ measurement will have negligible statistical uncertainties ($\sim$500 times smaller than at LEP, \ie\ amounting to about $\pm$0.005\%), and the experimental systematics will be reduced to the permille level thanks to an optimized control of the detector acceptances, efficiencies, and resolutions~\cite{FCCee}. The results of our study (Figs.~\ref{fig:lepton_AFBobs_vs_MC} and \ref{fig:jet_AFBobs_vs_MC}, and Table~\ref{tab:AFBbnew}) indicate that the jet-charge-based analyses have today QCD systematics uncertainties in the 3-permille range, at least twice better than the lepton-charge-based extraction. Even with the expected progress in the modeling of soft/collinear and higher-order QCD effects, the latter will still contribute to the systematics uncertainty budget of any $\AFBb$ measurement. In order to further reduce the impact of such QCD effects, it has been proposed to impose tighter acollinearity criteria between the reconstructed $b$ jets~\cite{AlcarazMaestre:2020fmp}. In any case, improvements of a factor of ten or more in our theoretical numerical control of $b$-, $c$-quark showering and hadronization and higher-order pQCD corrections ---as an integral part of a wider QCD-related experimental program~\cite{Anderle:2017qwx}--- will be needed to precisely measure electroweak observables such as $\AFBbb$ at any future $\epem$ collider running at the Z pole.

The latest parton shower algorithms developed ---such as \textsc{PanScales}~\cite{vanBeekveld:2023chs,vanBeekveld:2023ivn}, \textsc{Alaric}~\cite{Hoche:2024dee}, \textsc{Apollo}~\cite{Preuss:2024vyu},
\textsc{Deductor}~\cite{Nagy:2017dxh}, and \textsc{FHP}~\cite{Holguin:2020joq}--- describe the radiation of soft and collinear partons with at least one order higher logarithmic accuracy than the
MC event generators used to date. The upcoming incorporation of such next-to- (or next-to-next-to-) leading logarithmic developments into the latest MC event generators
(\pythia~8.3~\cite{Bierlich:2022pfr}, \herwig~7.2~\cite{Bellm:2019zci}, and \textsc{Sherpa}~3~\cite{Sherpa:2024mfk}), as well as the associated retuning of their hadronization settings,
will certainly improve the precision and accuracy of the theoretical tools needed to describe the past and upcoming $\epem$ hadronic data. Such improved QCD understanding will be key in the quest for
indirect signals of new physics through high-precision measurements of multiple SM observables at the FCC-ee~\cite{dEnterria:2019jfn,dEnterria:2025hbe}.
 
%%%%%%%%%%%%%%%%%%%%%%%%%%%%%%%%%%%%%%%%%%%%%%%%%%%%%%%%%%%%%%%%%%%%%%%%%%%%%%%%%%%%%%%%%%
\section{Summary and conclusions}
\label{sec:summary}

The forward-backward asymmetry ($\AFBbb$) of bottom quarks produced in $\epem\to\bbbar(g)$ processes, measured around $\sqrts = m_{\rm Z}$ at LEP, remains one of the few experimental measurements not fully consistent with the SM theoretical predictions. We have studied to what extent the QCD developments in the last twenty years have improved our understanding of the uncertainties related to hard gluon radiation, parton showering, and hadronization corrections in the original $\AFBbb$ experimental extractions. We implemented the eight original LEP $\AFBbb$ analyses ---four of them reconstructing the (leading) lepton-charge and four others the jet-charge--- in a \pythia\,8 simulation with nine different models of parton radiation and fragmentation, including the alternative \vincia\ shower model. %Within the comparatively large experimental uncertainties, we find an overall good agreement of our simulations with each one of the $\AFBobs$ measurements at LEP. 
The parton shower and hadronization uncertainties have been estimated from the spread of the $\AFBobs$ values derived from the nine different MC simulations. The total QCD uncertainties, including NNLO corrections, of the extracted asymmetries appear to be about a factor of two smaller than those of the original LEP results. The final propagated QCD uncertainties are 0.7\% and 0.3\% for the lepton- and jet- charge-based analyses, respectively. Naively combining all eight LEP measurements with their revised QCD uncertainties and with the massive $b$-quark NNLO theoretical pQCD corrections, would lead to a new $b$-quark FB asymmetry that changes slightly from $\AFBb =0.0992 \pm 0.0016$ to $0.0995\pm0.0016$, with a 0.3\% increase of its central value but basically unchanged uncertainties, dominated by statistical errors. This result would diminish the tension between the forward-backward $b$-quark asymmetry and the SM fit from 2.4$\sigma$ to 2.2$\sigma$. We have also shown that the QCD systematic uncertainties will ultimately impact the maximum precision attainable of any future $\AFBb$ measurement, and associated $\weakeffb$ determination, at future high-luminosity $\epem$ facilities running at the Z pole with orders of magnitude smaller statistical uncertainties, and have pointed out the importance of dedicated studies to minimize them.

\paragraph*{Acknowledgments--} Discussions with Peter Skands on \pythia\,8 and \vincia\ are acknowledged. The work of C.~Yan has been supported by the NSF (USA).

%%%%%%%%%%%%%%%%%%%%%%%%%%%%%%%%%%%%%%%%%%%%%%%%%%%%%%%%%%%%%%%%%%%%%%%%%%%%%%%%%%%%%%%%%%%%%%%%%
%\section*{References}

\end{document}